\documentclass{PoS2}
\usepackage[authoryear,square]{natbib}
\bibpunct{(}{)}{;}{a}{}{,}

\usepackage{multirow}

\newcommand{\ltsim}{\mbox{{\raisebox{-0.4ex}{$\stackrel{<}{{\scriptstyle\sim}}$}}}}

\title{Cosmology with SKA Radio Continuum Surveys}

               \ShortTitle{Cosmology with SKA Radio Continuum Surveys}

\author{
{Matt J.~Jarvis}$^{1,2}\thanks{Speaker}$, 
David Bacon$^3$, Chris Blake$^{4}$, Michael L.~Brown$^5$,Sam N.~Lindsay$^{1}$, Alvise Raccanelli$^{6,7,8}$, Mario Santos$^{2,9}$,
Dominik Schwarz$^{10}$
\\ 
$^1$Astrophysics, University of Oxford, Keble Road, Oxford, OX1
        3RH, UK; $^2$Physics Department, University of the Western
        Cape, Bellville 7535, South Africa; 
$^3$Institute of Cosmology and Gravitation, University of Portsmouth, Burnaby Road, Portsmouth PO1 3FX, UK;
$^4$ Centre for Astrophysics \& Supercomputing, Swinburne University of Technology, PO Box 218, Hawthorn, VIC 3122, Australia;
$^5$ Jodrell Bank Centre for Astrophysics, Alan Turing Building, University of Manchester, Oxford Road, Manchester, M13 9PL, UK; 
$^6$ Department of Physics \& Astronomy, Johns Hopkins University, 3400 N. Charles St., Baltimore, MD 21218, USA;
$^7$ Jet Propulsion Laboratory, California Institute of Technology, Pasadena CA 91109, USA;
$^{8}$ California Institute of Technology, Pasadena CA 91125, USA; 
$^9$SKA SA, 3rd Floor, The Park, Park Road, Pinelands, 7405, South Africa;
$^{10}$Fakult\"at fu\"ur Physik, Universit\"at Bielefeld, 33501 Bielefeld, Germany
\\
E-mail: \email{matt.jarvis@astro.ox.ac.uk}
}

\abstract{Radio continuum surveys have, in the past, been of restricted use in cosmology. Most studies have concentrated on cross-correlations with the cosmic microwave background to detect the integrated Sachs-Wolfe effect, due to the large sky areas that can be surveyed. As we move into the SKA era, radio continuum surveys will have sufficient source density and sky area to play a major role in cosmology on the largest scales. In this chapter we summarise the experiments that can be carried out with the SKA as it is built up through the coming decade. We show that the SKA can play a unique role in constraining the non-Gaussianity parameter to $\sigma_{\rm f_{NL}} \sim 1$, and provide a unique handle on the systematics that inhibit weak lensing surveys. The SKA will also provide the necessary data to test the isotropy of the Universe at redshifts of order unity and thus evaluate the robustness of the cosmological principle.
Thus, SKA continuum surveys will turn radio observations into a central probe of cosmological research in the coming decades. 
}

\FullConference{
Advancing Astrophysics with the Square Kilometre Array\\
June 8-13, 2014\\
Giardini Naxos, Italy}

\begin{document}

\section{Introduction}

\noindent
Over the past decade it has become clear that large-area radio surveys can play an important role in enhancing our understanding of the cosmological model. Various groups have used the NRAO VLA Sky Survey \citep[NVSS;][]{Condon1998} as a low-redshift tracer of the large-scale structure of the Universe by both cross-correlations with the Cosmic Microwave Background (CMB) to find evidence for the Integrated Sachs-Wolfe effect \citep{SachsWolfe1967,Giannantonio2008}, and by investigating the largest scales for signs of non-Gaussianity \citep[e.g.][]{Xia2010}.

As we move into the SKA-era, then such wide-area surveys will reach much deeper levels, and begin to rival optical surveys in terms of source density. This is because the synchrotron emission that dominates the extragalactic radio background emission at $<10$~GHz is emitted from all galaxies with any ongoing star formation or accretion activity. In addition, the wide area radio surveys will extend to substantially larger areas ($\simeq 30,000$~deg$^2$) than many of the forthcoming optical surveys.

Given that synchrotron radio emission from galaxies is unaffected by dust obscuration, and the $k-$corrections are relatively straightforward given the power-law nature of the synchrotron spectrum, then radio surveys offer a unique advantage for many cosmological applications. On the other hand, the generally featureless spectrum means that redshift information for individual sources can only be obtained with observations at other wavelengths, or through 21-cm spectral line observations \citep[see e.g.][]{Bull2014}.

Like optical surveys,  radio surveys are made up of sources that trace the underlying density field very differently, i.e. from highly biased tracers such as powerful AGN to low-bias populations such as star-forming galaxies. However, unlike the wide-field optical surveys, radio continuum emission is unaffected by dust, and in the age of the SKA, the star-forming galaxies, as well as the AGN, can be detected to high redshifts. These sources can then provide an important probe of the very largest scales in the universe.

In this chapter we review the different experiments and surveys that could be carried out with the SKA in phase 1, and in the longer term with the full SKA, for cosmology using radio continuum observations.

\section{Assumptions}

\noindent
We  base our estimates of the redshift distribution of various source populations on the luminosity functions that underpin the semi-empirical extragalactic
sky simulations of \cite{Wilman2008,Wilman2010}. These simulations provide a very good description of the latest
source counts from various deep field surveys with the JVLA
\citep[e.g.][]{Condon2012}. Although modifications may be required
to accurately reproduce the most recent results from e.g. {\em Herschel},
the general trends and evolution prescribed are relatively well
matched to our current understanding, and the extrapolations to
flux-density levels yet to be reached in the radio band are
constrained by observations at a range of other wavelengths. For
example, in the chapter on weak lensing at radio wavelengths
and testing the foundations of cosmology \cite{Brown2014} and \cite{Schwarz2014}, both use an increase in the source density of the star-forming
population by a factor of 2.5, in order to match the latest source
counts from deep fields. In Table~1 we present the number
count for each radio source population for four flux-density
thresholds at $\nu = 1$~GHz, which are representative of the typical surveys that could
be conducted with the SKA1 and SKA2. However, we note that moving the frequency to around 700~MHz would lead to an increase in the field-of-view by factor of $\sim 2$ and an increase in the source density that scales with the spectral index, i.e. $S_{\nu} \propto \nu^{\alpha}$, with $\alpha = -0.7$ for a typical star-forming galaxy.

\begin{table}
\caption{Expected number density (deg$^{-2}$) of the various
  radio source populations in the simulations of
  \cite{Wilman2008} for a survey
  with detection threshold of $S_{\rm 1~GHz}$ = 100~nJy ({\em top-left}), 1$\mu$Jy ({\em
    top-right}), 5$\mu$Jy ({\em
    bottom-left}) and 10$\mu$Jy ({\em
    bottom-right}). SFGs=star-forming galaxies; SBs=starburst
  galaxies; RQQs=radio-quiet quasars; FRIs=Fanaroff-Riley Class I
  sources; FRIIs=Fanaroff-Riley Class II sources. Note that the number density of SFGs and SBs in the table
may be multiplied by a factor of 2.5 to match the latest
observed source
counts at the $\mu$Jy level  \citep[see][ for further details]{Brown2014}.}\label{tab:sims}
{\scriptsize
\begin{tabular}{|c|ccccc|c|ccccc|c|}
\hline
& \multicolumn{6}{c|}{ 100~nJy } & \multicolumn{6}{c|}{  1~$\mu$Jy }\\
\hline
Redshift & \multicolumn{6}{c|}{ $N$/deg$^2$ } & \multicolumn{6}{c|}{ $N$/deg$^2$ }\\
& SFGs & SBs & RQQs & FRIs & FRIIs & Total &  SFGs & SBs & RQQs & FRIs & FRIIs & Total  \\
\hline
0.0$<z<$0.5 &   5892  &     35  &   4530  &    358 & 0.0 & 10815 &   3419  &     37  &   1710  &    279 & 0.0 & 5445 \\
0.5$<z<$1.0 &  15366  &    191  &   7093  &   2833 & 0.2 & 25483&   7759  &    260  &   1964  &   1004 & 0.2 & 10987\\
1.0$<z<$1.5 &  19353  &    731  &   4311  &   3191 & 0.8 &  27587 & 9181  &    666  &   1466  &    857 & 0.8 & 12171\\
1.5$<z<$2.0 &  18102  &    599  &   2478  &   1920 & 1.2 &  23100 & 8042  &    719  &    967  &    598 & 1.2 & 10327 \\
2.0$<z<$2.5 &  16455  &   1103  &   1828  &   1706 & 1.0 &21093  & 5844  &    697  &    664  &    449 & 1.0 & 7655\\
2.5$<z<$3.0 &  13572  &    888  &   1104  &   1198 & 1.6 & 16764  & 4141  &    525  &    438  &    328 & 1.6 & 5434\\
3.0$<z<$3.5 &  11582  &    925  &    787  &    895 & 0.8 & 14190 & 2970  &    369  &    281  &    249 & 0.8 &3870\\
3.5$<z<$4.0 &  10298  &    887  &    613  &    848 & 0.3 & 12646&  2213  &    302  &    204  &    209 & 0.3 &2928 \\
4.0$<z<$4.5 &   8144  &    790  &    414  &    579 & 0.3 & 9927 & 1676  &    232  &    145  &    163 & 0.3 &2216 \\
4.5$<z<$5.0 &   7026  &    649  &    306  &    481 & 0.2 & 8462  & 1259  &    184  &    102  &    130 & 0.2 &1675\\
5.0$<z<$5.5 &   6037  &    720  &    233  &    415 & 0.0 & 7405&  1005  &    154  &     75  &    110 & 0.0 &1344\\
5.5$<z<$6.0 &   5257  &    638  &    182  &    319 & 0.0 & 6396  &  803  &    126  &     54  &     93 & 0.0 &1076\\
\hline
Total & 137084 & 8156  & 23879 & 14743& 6.4& 183868 & 48312 & 4271 & 8070 & 4469 & 6.4 & 65128\\
\hline
\hline
& \multicolumn{6}{c|}{ 5~$\mu$Jy } & \multicolumn{6}{c|}{
  10~$\mu$Jy }\\
\hline
Redshift & \multicolumn{6}{c|}{ $N$/deg$^2$ } & \multicolumn{6}{c|}{ $N$/deg$^2$ }\\
& SFGs & SBs & RQQs & FRIs & FRIIs & Total & SFGs & SBs & RQQs & FRIs & FRIIs & Total \\
\hline
0.0$<z<$0.5 &   1761  &     40  &    701  &    140 & 0.0 & 2642 &  1117  &     40  &    445  &     99 & 0.0 & 1701 \\
0.5$<z<$1.0 &   2894  &    267  &    913  &    425 & 0.2 & 4499 &  1594  &    208  &    636  &    293 & 0.2 & 2731 \\
1.0$<z<$1.5 &   2973  &    368  &    713  &    351 & 0.5 & 4405 &   1588  &    225  &    495  &    239 & 0.5 & 2548 \\
1.5$<z<$2.0 &   2350  &    317  &    468  &    245 & 1.2 & 3381 &   1206  &    192  &    309  &    167 & 1.2 & 1875 \\
2.0$<z<$2.5 &   1462  &    218  &    290  &    184 & 1.0 & 2155 &   693  &    129  &    179  &    124 & 1.0 & 1126 \\
2.5$<z<$3.0 &    930  &    162  &    174  &    136 & 1.2 & 1403 &   405  &     95  &     97  &     92 & 1.2 & 690 \\
3.0$<z<$3.5 &    605  &    118  &    103  &    105 & 0.7 & 932 &   239  &     67  &     53  &     71 & 0.7 & 431 \\
3.5$<z<$4.0 &    407  &     91  &     66  &     84 & 0.3 & 648 &   145  &     50  &     31  &     56 & 0.3 & 282 \\
4.0$<z<$4.5 &    274  &     69  &     39  &     66 & 0.4 & 448 &    87  &     38  &     17  &     44 & 0.4 & 186 \\
4.5$<z<$5.0 &    185  &     53  &     26  &     53 & 0.1 & 317 &    53  &     28  &      9  &     35 & 0.1 & 125 \\
5.0$<z<$5.5 &    130  &     42  &     17  &     42 & 0.0 & 231 &    34  &     21  &      5  &     28 & 0.0 & 88 \\
5.5$<z<$6.0 &     91  &     35  &     10  &     36 & 0.0 & 172 &    21  &     17  &      3  &     24 & 0.0 & 65\\
\hline
Total  & 14062 & 1780 & 3520 & 1867 & 5.6 & 21235 & 7182 & 1110 & 2279 & 1272 & 5.6 & 11849 \\
\hline
\end{tabular}
}

\end{table}

\section{Evolution of bias for radio sources}\label{sec:bias}

\noindent
One of the key measurements that can be extracted from surveys is knowledge of the bias of the population, i.e. how the light emitted traces the underlying dark-matter distribution. It will be very challenging to measure the evolution of the bias of radio sources from the large-scale surveys that will predominantly be used for measuring the cosmological signal, due to the lack of redshift information. However, assuming that the bias of a given source population is scale-invariant, then deeper surveys with a greater degree of multi-wavelength information can be used to measure both the redshift distribution and the bias of the radio sources, in conjunction with measuring the large-scale structure for the wide-area surveys. Such data will also be important for understanding the effects that baryons have on the underlying dark matter, and how this in turn affects  the clustering signal.

\begin{figure*}
\includegraphics[width=15cm]{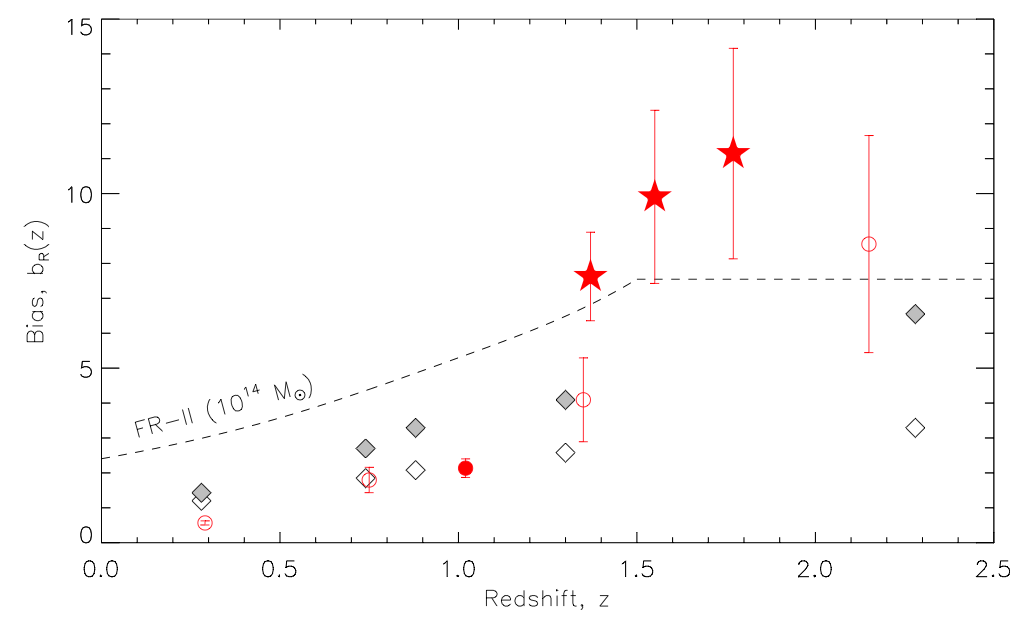}
\caption{Linear bias of near-infrared identified radio sources as a function of median redshift. Open circles correspond to the four independent redshift bins used while the filled circle is the bias for the full sample of 766 radio sources with $Ks < 23.5$ and $S_{1.4} > 90\mu$Jy. Star symbols correspond to lower luminosity limits of $10^{23}$, $10^{23.5}$ and $10^{24}$~W~Hz$^{-1}$ from low to high redshift. The dashed line shows the FR-II bias adopted by \citep{Wilman2008} in the SKADS simulations, and the diamond symbols show the expected bias based on the SKADS prescriptions (open) and with the FR-I halo mass increased to $10^{14}$~M$_{\odot}$ (filled), matching the FR-IIs. Taken from \cite{Lindsay2014b}.}\label{fig:bias}
\end{figure*}

An initial attempt at constraining the bias of the radio source population was carried out by \cite{Lindsay2014b,Lindsay2014a} who used a combination of wide-shallow and deep-narrow radio and optical/near-infrared survey data to determine the clustering of the radio source population in redshift shells (Figure~\ref{fig:bias}). They found that the dominant AGN population, namely the low-luminosity (FRI-type sources as defined in the SKADS simulations) appear to have a higher bias than currently simulated. This provides an opportunity to use the multi-tracer technique to overcome the debilitating effects of cosmic variance on large-scale measurements of the power spectrum using various probes \citep{Seljak2009}. 

However, such work is not limited to small scales, where the ancillary data provides the radial information. With the SKA we will also be in a position to cross-correlate the radio source population with CMB-lensing maps from current \citep[e.g.][]{PlanckLens,ACTLens} and future CMB facilities. Some work along these lines has already been performed  \citep{Sherwin2012,Geach2013}, providing a unique measurement of the bias for high-redshift quasars. 

With information about the redshift distribution of the various sub-populations of radio sources (e.g. star-forming/AGN type) in any deep radio survey, then this information can inform the measurement of the bias from cross-correlating with CMB-lensing maps. This can provide a robust measurement of both the weighted bias of all populations, or if they can be morphologically distinguished \citep[see][]{Makhatini2014}, the bias and its evolution for each individual source population.

\section{Cosmological Probes}

\subsection{The Angular Power Spectrum}\label{sec:powerspectrum}

\noindent
In the the absence of redshift information, the most straightforward experiment is to measure  angular correlation function or power spectrum of the radio sources. \cite{Raccanelli2012} and \cite{Camera2012} have already shown that a measurement of the angular power spectrum, in combination with the CMB and supernovae Ia, can be a very useful tool to determine both the dark energy equation of state and departures from general relativity, even with the precursor surveys within reach of ASKAP, LOFAR and MeerKAT. However, to make radio surveys competitive with the largest surveys at other wavelengths then we require the combination of source density, sky area and morphological characterisation that is only feasible with the SKA.

\begin{figure*}
\includegraphics[width=15cm]{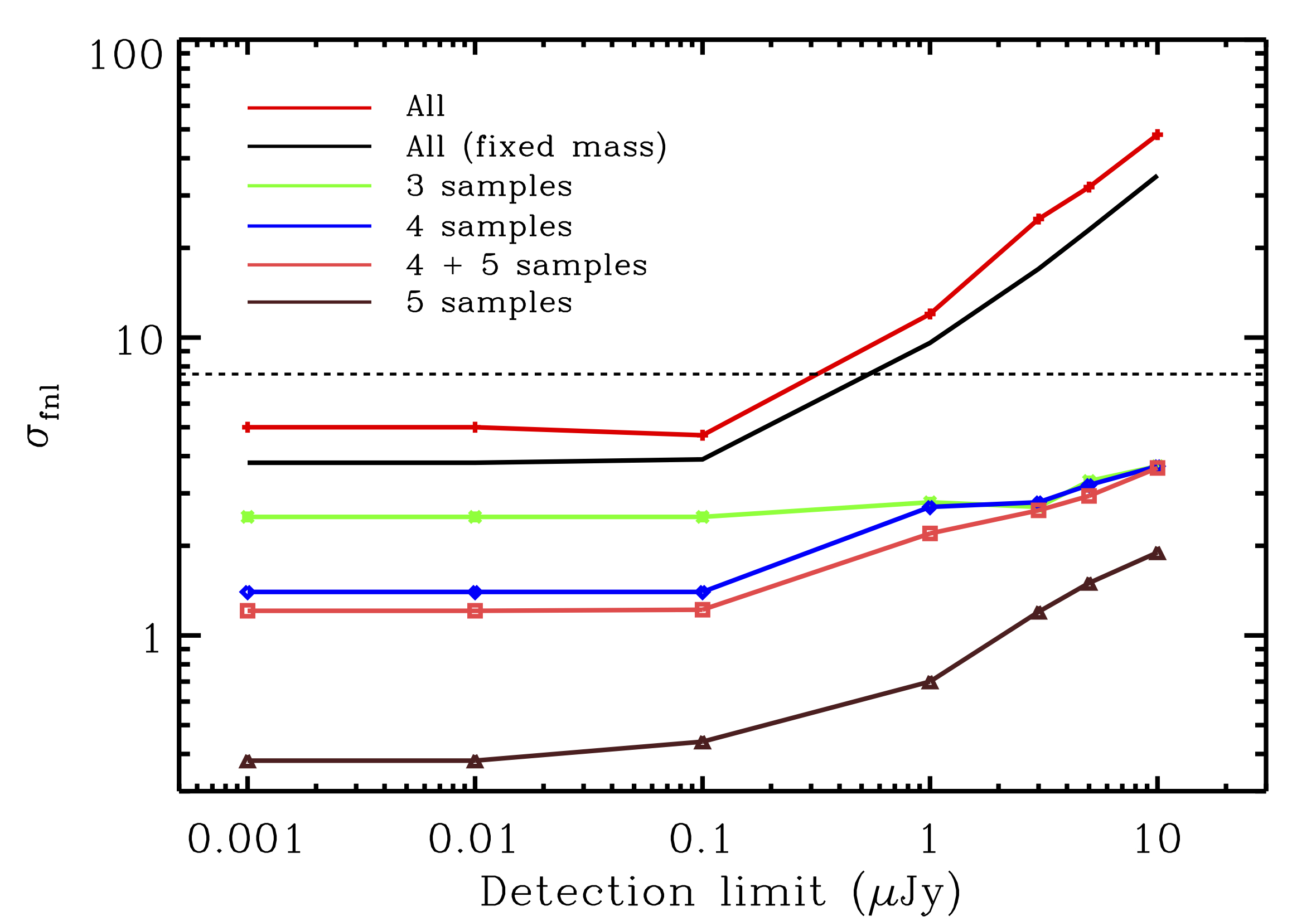}
\caption{Forecast constraints on $f_{NL}$ obtained with the multi-tracer method as a function of the flux-density threshold used to detect galaxies. The various populations considered are FR-I and FR-II radio galaxies, radio-quiet quasars, star-forming galaxies and starburst galaxies, with different biases, as described in both \cite{Wilman2008} and \cite{Ferramacho2014}. 
We present the results obtained using the full sample of objects with an averaged effective bias and those obtained using the combination of 3 populations of radio galaxies (where SRG, SB and RQQ correspond to one population group), using 4 populations (where only SFG and SB are undifferentiated) and with a selection of 5 populations for $z<1$ and 4 populations for $z> 1$ (again with undifferentiated SFG and SB). We also show the result for the ideal case where all 5 populations could be differentiated over the entire redshift range of the survey. The horizontal line represents the best constraint obtained by the Planck collaboration \citep{PlanckNG}. Taken from \cite{Ferramacho2014}. }\label{fig:NG}
\end{figure*}

Recently, \cite{Ferramacho2014} demonstrated that even without redshift information for the individual sources, or a subset of sources, the wide-area radio continuum surveys can play a unique role in constraining the level of non-Gaussianity. Utilising the multi-tracer technique \citep{Seljak2009}, they showed that the different populations of radio sources, which trace the underlying dark-matter distribution with vastly different biases, can constrain the local non-Gaussian parameter $f_{NL}$ with uncertainty $\sigma_{\rm f_{NL}}=3.6$  for a galaxy detection flux limit of 10~$\mu$Jy and $\sigma_{\rm f_{NL}}=2.2$ for 1~$\mu$Jy (Figure~\ref{fig:NG}). The former survey is within reach of SKA1, but requires good resolution in order to morphologically distinguish the different classes of radio source, i.e. FRI/FRII from star-forming galaxies and radio-quiet quasars. As shown in \cite{Makhatini2014} this sort of classification is possible to very high redshift with SKA1-MID. 

Therefore radio surveys with SKA1-MID, without any additional data, have the potential to constrain primordial non-Gaussianity to a factor of $\sim 2$ better than the present constraints obtained with Planck. This leads to the possibility of obtaining $\sigma_{\rm f_{NL}}=2.2$ with SKA2 (Figure~\ref{fig:NG}).

Furthermore, the late time ISW effect allows us to strongly constrain a class of models that can have $w_0$ and $w_a$ very similar to a quintessence model, or even to $\Lambda$CDM, but that allows for viscous pressure. The set of models
that are called unified dark sector models or similar (e.g. a 
single viscous dark fluid) could in principle resemble LCDM in $w_a$ and
$w_0$, and produce a much larger ISW effect \citep[see e.g.][]{LiBarrow2009,Velten2011}.

\subsection{Cosmic Dipole, Isotropy and Homogeneity}\label{sec:homogeneity}

All-sky continuum surveys with the SKA will probe large angular scales at a median redshift 
of order unity. This will allow tests of fundamental assumptions of modern cosmology, 
especially of statistical isotropy. Combined with spectroscopic redshifts, either from SKA HI 
surveys, or by means of other instruments, statistical homogeneity of the cosmos will be tested.
SKA will allow us to study scales at $z \sim 1$ that have not been in causal contact since 
the first horizon crossing during inflation, and therefore contain information that was 
frozen in during cosmological inflation. 

\begin{figure*}
\includegraphics[width=5.4cm,angle=270]{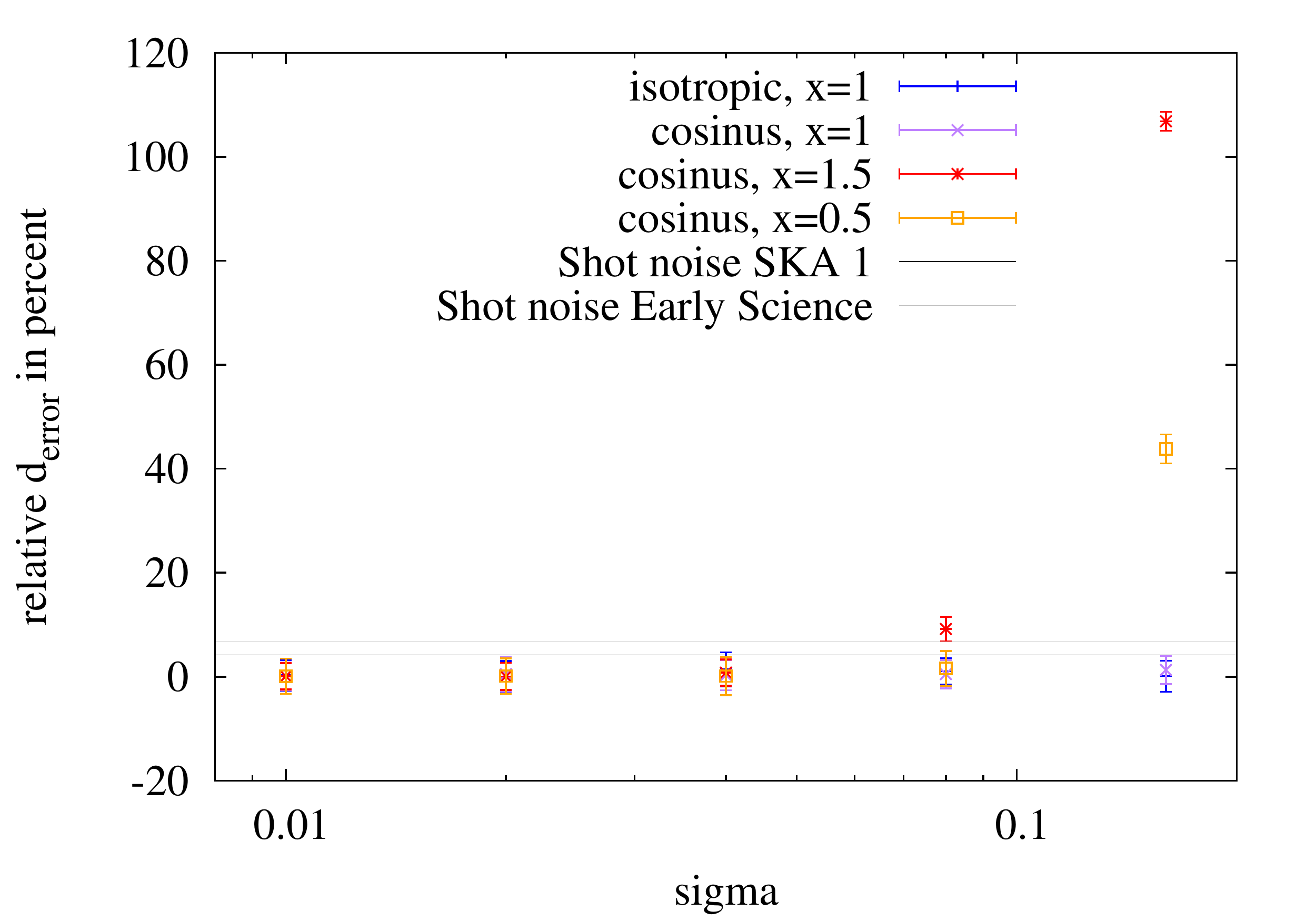}
\includegraphics[width=5.4cm,angle=270]{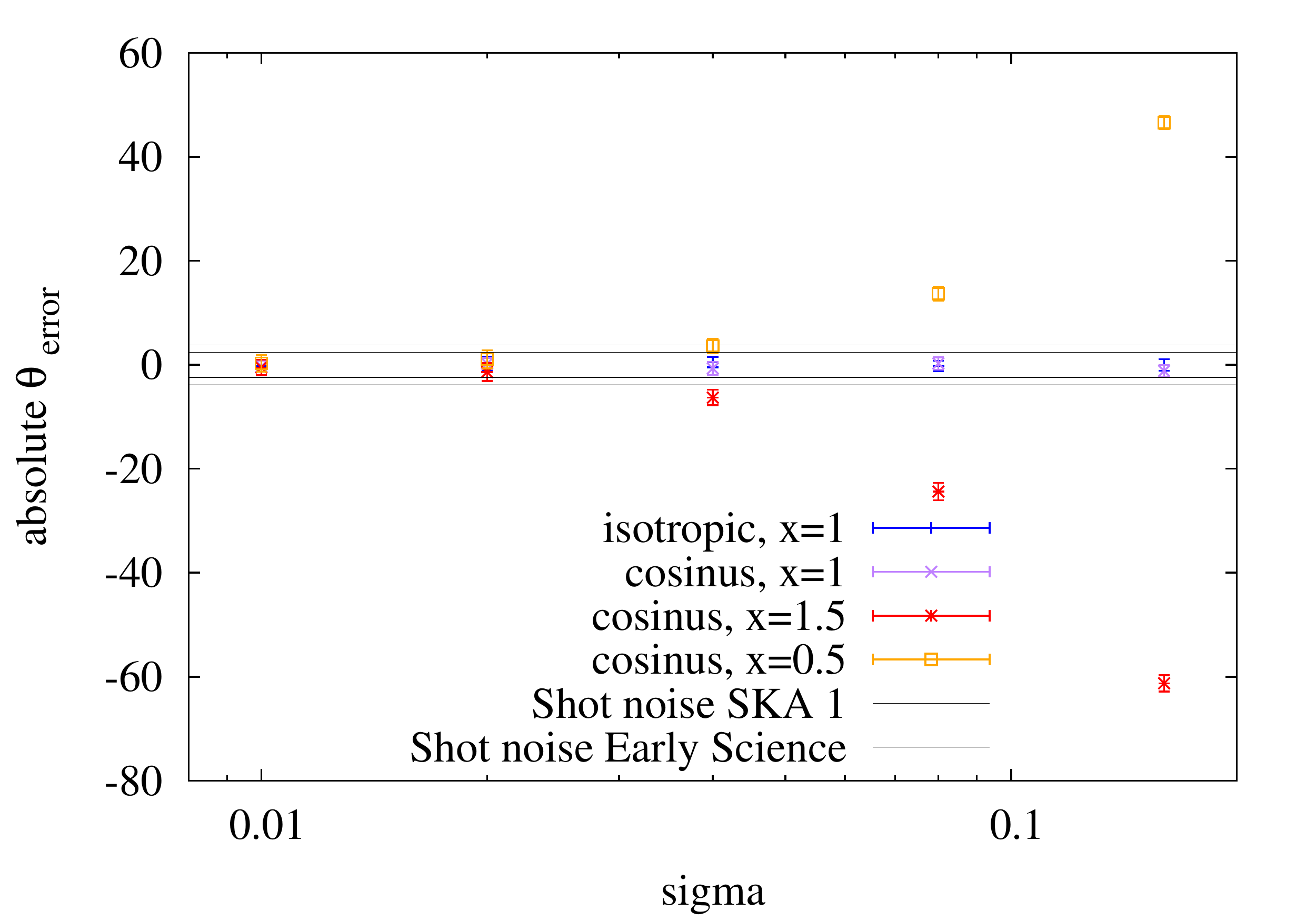}
\caption{Left: Accuracy (in per cent) of the measurement of the dipole amplitude as function of 
fractional error on flux density calibration on individual point sources (sigma). All points are based on 
100 simulations. Right: Accuracy (in degrees) of the measurement of the dipole direction. The 
horizontal lines denote the error due to shot noise for a dipole
estimate based on $8 \times 10^7$ sources 
(SKA Early Science) and $2\times 10^8$ sources (SKA1). See \cite{Schwarz2014} for further details.}\label{fig:isotropy}
\end{figure*}

SKA all-sky surveys will allow the measurement of the cosmic radio dipole almost as precisely as 
the CMB dipole. SKA1 will constrain the cosmic radio dipole direction with an accuracy better 
than 5~degrees (Fig.~\ref{fig:isotropy}), and SKA2 within a degree (at 99 per cent C.L.). 
Compared to todays best estimate based on NVSS data, this will be an improvement of a factor of 
100 in the accuracy of the cosmic radio dipole direction for SKA1. This measurement could 
firmly establish or refute the commonly adopted assumption that the CMB and the overall 
large-scale structure frames agree. 

The CMB exhibits unexpected features at the largest angular scales, among them a lack of angular 
correlation, alignments between the dipole, quadrupole and octopole, hemispherical asymmetry, a 
dipolar power modulation, and parity asymmetries \citep{PlanckIsotropy,Copi2013a,Copi2013b}. 
Understanding the statistical significance of these anomalies is crucial, as a lack of statistical 
isotropy or Gaussianity could rule out the standard cosmological model. 

The precision of these CMB measurements is limited by our understanding of the foregrounds, and 
observational uncertainties are already much smaller than the cosmic variance at such scales. 
Therefore, it is very difficult to identify the cause of these anomalies without an independent 
probe at the same scales.

The angular two-point correlation at angles $> 60$ degrees from SKA continuum surveys, as well 
as the reconstruction and cross-correlation of low multipoles will offer further insight into 
these puzzles, see \cite{Schwarz2014}.

Supplemented with spectroscopic information, several tests of homogeneity can be envisaged. One 
possibility is to probe a consistency relation of Friedmann-Lema\^itre models, which involves 
the dimensionless comoving distance, the Hubble rate, and their derivatives \citep{Clarkson2008}.  

Thus, studying the large-angular scales in SKA continuum surveys might help resolve the puzzle
of CMB anomalies and test the cosmological principle.

\subsection{BAO and RSDs with spectroscopic follow-up of emission-line sources }

\noindent
Although not a primary case for the SKA, radio continuum sources are, by definition, active and exhibit emission lines in their optical spectra. This property can be utilised to perform efficient follow-up spectroscopy with the next generation of optical and near-infrared spectrographs, in much the same way that the WiggleZ survey targeted UV-bright sources \citep[e.g.][]{Blake2012}. Such follow-up would provide an efficient method for extracting precise redshift information for a very large number of sources over a large fraction of the sky. This could provide new and robust measurements of the Baryon Acoustic Oscillations up to high-redshift (potentially targeting emission line objects at $z>2$), as well as redshift-space distortions.

Rather than a key driver for the SKA, this demonstrates a unique synergy, and something that is planned to be tested by combining the LOFAR extragalactic surveys with the new multi-object spectrographs (MOS) that are currently in development (see Section~\ref{sec:multiwavelength}).


\subsection{Integrated Sachs-Wolfe Effect}

\begin{figure*}
\includegraphics[width=15cm]{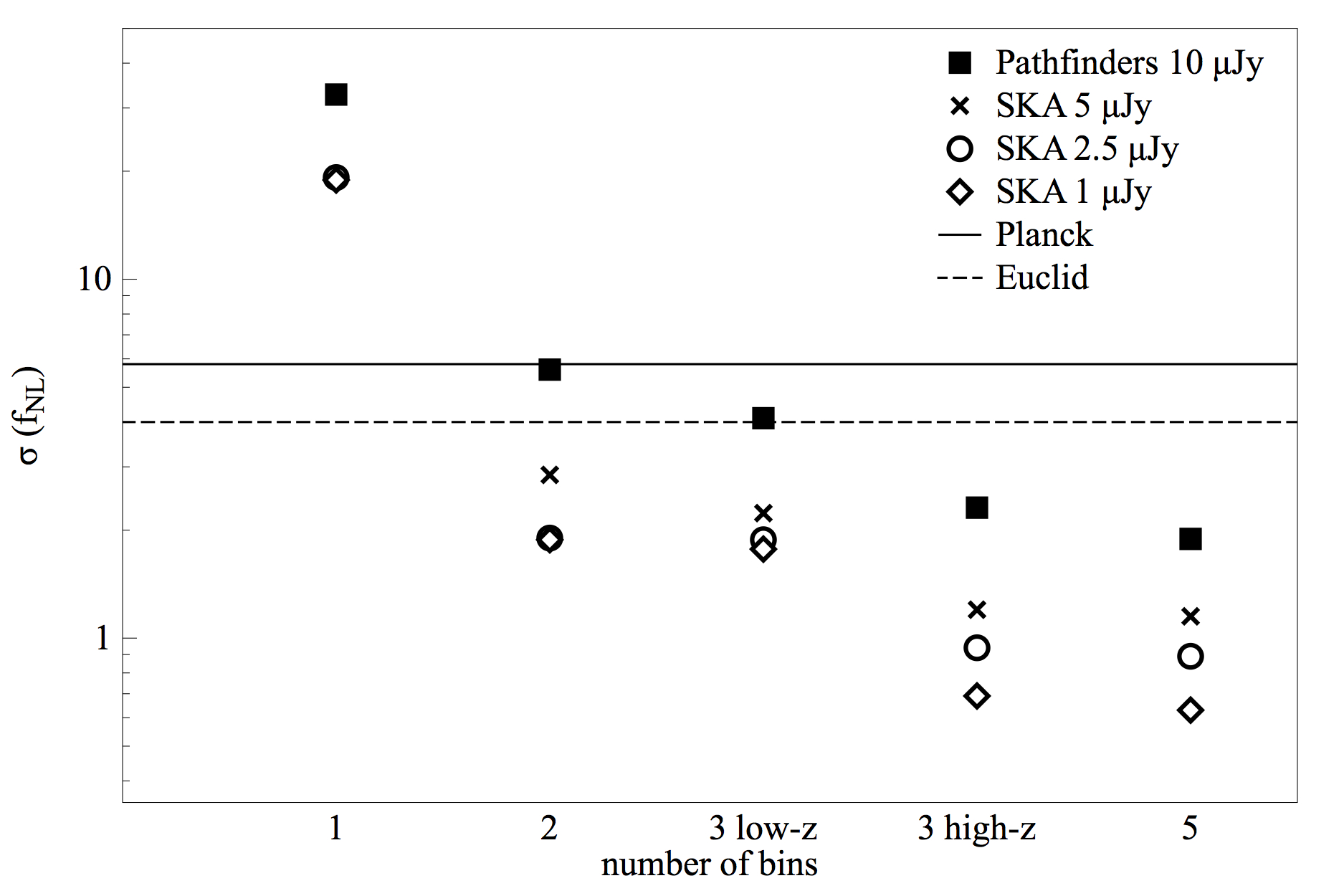}
\caption{Constraints on ($f_{\rm NL}$) with integrated Sachs-Wolfe measurements using SKA continuum surveys. The effect of having several redshift bins for the measurements of the $f_{\rm NL}$ parameter for different flux-density limits that will be possible for various stages of the SKA. For full details see \cite{Raccanelli2014}.}\label{fig:ISW}
\end{figure*}

\noindent The ISW effect \citep{SachsWolfe1967} manifests itself in the correlation between large-scale structure and CMB temperature. In an Einstein-de Sitter universe, the energy gained by a photon falling into a gravitational potential well is exactly cancelled out by the energy lost upon climbing out of the well. In a universe with a dark energy component or modified gravity to the same effect, the local gravitational potential varies with time, and potential wells are stretched throughout the crossing time of the photons. This disparity between potential on entry and exit imparts a net blueshift on the incident photon and, equivalently, an increase in photon temperature. Specifically, this effect is important at late times ($z \ltsim 1$), once the Universe had begun its transition towards a dominant dark energy component and accelerated expansion.
The ISW signal is small, compared to the intrinsic temperature anisotropies in the CMB, acting on large scales where cosmic variance most affects CMB uncertainties, and so cross-correlation with local large-scale structure, with extensive sky coverage, is required in order to produce a significant result. \cite{Giannantonio2012} discuss the current state of ISW measurements, and potential problems. While several detections of the ISW effect had been made previously, cross-correlating the CMB maps from the Wilkinson Microwave Anisotropy Probe (WMAP) with radio, infrared, optical, and X-ray surveys \cite[for results using radio surveys see e.g.][]{Nolta2004,Raccanelli2008}, these were all at rather low significance. \cite{Giannantonio2008} reached an increased significance of $\sim 4.5\sigma$ by combining surveys to develop a fuller catalogue of local large-scale structure. With the latest CMB measurements from the Planck satellite, and corresponding all-sky surveys of massive, low-redshift galaxies (and an understanding of their relationship to the underlying dark matter distribution), the ISW could be a powerful cosmological tool \citep[Figure~\ref{fig:ISW};][]{Raccanelli2014}.

\subsection{Cosmic Magnification Bias}

\noindent The presence of mass alters the geodesic followed by light rays, causing the deflection of those rays along the line of sight when passing by intervening large-scale structure. This causes distortions and magnification in the observed images of distant astronomical sources, and is known as gravitational lensing. The shape distortion of the background galaxies (`cosmic shear') can be used to constrain some cosmological parameters (see below, section~\ref{sec:weaklensing}), as can the magnification of background sources by foreground lensing structures (`cosmic magnification'). 

This magnification is governed by two effects. The increased flux received from distant sources due to lensing has the effect of bringing into the survey sources which would otherwise have been too faint to be detected. The lensing also stretches the solid angle, reducing the apparent surface density of the lensed background sources. Gravitational lensing should, therefore, leave a cosmic magnification signal in the angular cross-correlation function of two samples of sources with non-overlapping redshift distributions, since the foreground sample lenses the background sample. This signal, the magnification bias, is dependent on the balance struck between the loss of sources due to dilution, and the gain due to flux magnification.

Using the galaxy-quasar cross-correlation function, Scranton et al. (2005) made an $8 \sigma$ detection of the cosmic magnification signal of quasars lensed by foreground galaxies, both selected from the Sloan Digital Sky Survey (SDSS). They found that bright quasars, with steep source counts, exhibited an excess around foreground structure, and faint quasars, with shallow source counts were in deficit. Since this first success, Hildebrandt, van Waerbeke \& Erben (2009) have made detections with normal galaxies from the Canada-France-Hawaii Telescope Legacy Survey (CFHTLS), and Wang et al. (2011) have done likewise in the far-infrared with {\em Herschel}. 

\cite{Raccanelli2012} have shown that the cosmic magnification signal in SKA pathfinders can contribute to cosmological constraints; the cross-correlations envisioned there are between radio surveys (for the background sample) and optical surveys such as SDSS (for the foreground sample, and removal of radio foreground). With SKA1 and SKA2, the continuum survey can provide the background sample, while the HI survey can provide the foreground sample and removal of radio foreground (in combination with optical surveys such as Euclid and/or LSST). This will allow us to provide constraints on bias, dark energy and gravity parameters, with a method quite different from lensing autocorrelation of source shapes; this affords an important check on systematic effects which could enter into shape measurement.

\subsection{Weak Lensing}\label{sec:weaklensing}

\noindent Weak gravitational shear is the coherent distortion in the shapes of
distant galaxies due to the deflection of light rays by intervening
mass distributions. Measurements of the effect on large scales is
termed ``cosmic shear'' and has emerged as a powerful probe of
late-time cosmology over the last 15 years
\citep[e.g.][]{Heymans2012}. Since gravitational lensing is sensitive to
the \emph{total} (i.e. dark plus baryonic) matter content of the
Universe, it has great potential as a very robust cosmological probe,
to a large degree insensitive to the complications of galaxy formation
and galaxy bias. One of the most promising aspect of weak lensing
measurements is their combination with redshift information: such
measurements are then a sensitive probe of both the geometry of the
Universe and the evolution of structure over the course of cosmic
time. In turn, these latter effects are dependent on the nature of the
dominant dark energy component in the Universe and/or on modifications
to the theory of General Relativity on large scales.

\begin{figure*}
\includegraphics[width=15cm]{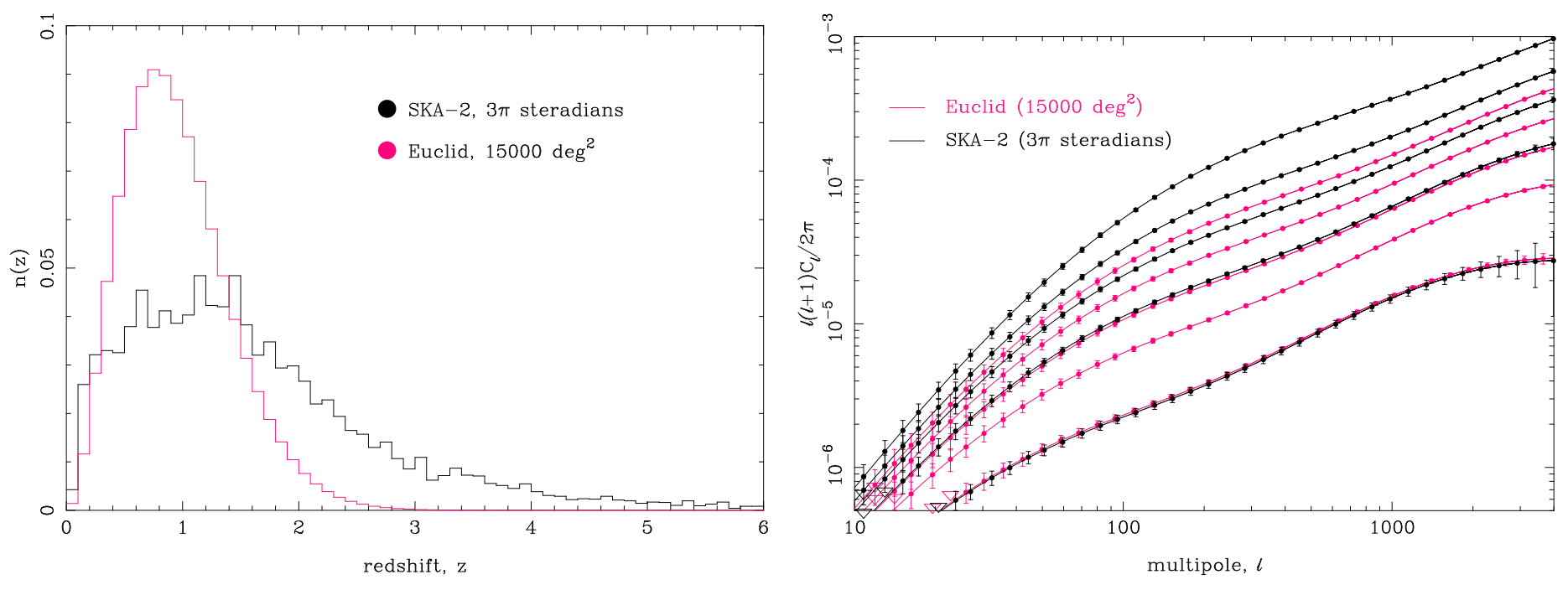}
\caption{Left panel: The redshift distribution of source galaxies for a $3\pi$ steradian weak lensing survey with the full SKA. Also shown is the redshift distribution for the 15,000~deg$^2$ survey with $Euclid$ . The n(z) extends to higher redshifts in the radio survey and probes a greater range of cosmic history. Right panel: The corresponding constraints on a 5-bin tomographic power spectrum analysis. For both experiments, an RMS dispersion in ellipticity measurements of $\gamma_{\rm rms} = 0.3$ is assumed, and the tomographic bins have been chosen such that the bins are populated with equal numbers of galaxies. Open triangles denote 1$\sigma$ upper limits on a bandpower. Note that only the auto power spectra in each bin are displayed though much cosmological information will also be encoded in the cross-correlation spectra between the different $z-$bins. See \cite{Brown2014} for more information.}\label{fig:lensing}
\end{figure*}

To date, the field of weak lensing has largely been the preserve of
optical surveys due to the much larger number densities of background
galaxies achieved in such surveys. However, this will change with the
advent of the Square Kilometre Array, which will reach galaxy number
densities of up to $\sim10$ galaxies arcmin$^{-2}$ with SKA1 and
$\sim75$ galaxies arcmin$^{-2}$ with SKA2 (Figure~\ref{fig:lensing}).  In addition, as described
in the weak lensing chapter in this volume \citep{Brown2014}, the radio offers truly unique 
approaches to measuring weak lensing that are (i) not available to optical 
surveys and (ii) potentially extremely powerful in minimising the most worrying
systematic effects in weak lensing cosmology. In particular, the PSF of an interferometer is completely deterministic (at least at the mid frequencies of the SKA), and the intrinsic synchrotron spectrum is smooth and featureless, thus the colour-dependent PSF should be far more straightforward to account for, cf. broad-band optical imaging. Moreover, using complementary radio shape information, along with polarisation information, which will be measured by default in the SKA continuum surveys, will help to control shape measurement systematics and the impact of intrinsic alignments. For SKA1 the radio continuum surveys will therefore be very complementary to the next generation of optical weak-lensing surveys from both the ground (e.g. DES)  and space (e.g. {\em Euclid}).

In the longer term, surveys with SKA2 will extend the reach of weak lensing beyond that of the optical mega-surveys of LSST and {\em Euclid}. First, the
typical redshifts probed by SKA surveys will go beyond that of optical surveys. In addition, the size of survey which SKA will be capable of achieving will by $\simeq 30,000$~deg$^2$, twice that of Euclid and only rivalled (or complemented) by LSST (see Figure~\ref{fig:lensing}).

However, as highlighted in \citep[e.g.][]{Patel2014} shape measurement at radio wavelengths need a similar effort to that spent in extracting shear estimates from ground-based optical surveys.

\section{The multi-wavelength requirements}\label{sec:multiwavelength}

\subsection{Spectroscopic redshifts}

\noindent
As with most cosmological measurements, greater accuracy can be achieved if higher precision redshifts can be obtained. In particular, using radio sources as targets for next generation MOS facilities could make such surveys highly efficient due to the higher likelihood of obtaining redshifts from emission lines.

As detailed in \cite{Jarvis2014} the future MOS facilities such as MOONS \citep{CirasuoloMOONS2012,MOONS2}, Prime Focus Spectrograph \citep[PFS; ][]{TakadaPFS2014} and the Maunakea Spectroscopic Explore \footnote{http://mse.cfht.hawaii.edu} (MSE) will provide a very good complement to SKA surveys, albeit only over around half of the SKA sky for both PFS and MSE due to them being sited in Hawaii.
For the wider surveys, 4MOST \citep{4MOST} would provide a basis for obtaining redshifts for the brighter star-forming galaxies, predominantly in the low-redshift Universe.

\subsection{Photometric redshifts}

\noindent
The bulk of the radio sources detected at these faint levels will be too faint at optical wavelengths to obtain spectroscopic redshifts. We will therefore be reliant on photometric redshifts. In the early phases this will be from surveys that are currently underway \citep[e.g.][]{McCracken2012,Jarvis2013,KIDS,VIKING,Banerji2014}

As we move to the full operation of SKA1 then we should also have LSST \citep[see ][]{Bacon2014} and {\em Euclid} \citep[see ][]{Kitching2014}, which will provide very deep imaging from the g-band through to H-band across a large swathe of the southern sky.

\section{Survey requirements for SKA1}

\subsection{Deep}

\noindent Although the main cosmological science will come from large-area surveys with the SKA, it is also important to conduct a deep survey with the best ancillary data available. As outlined in various chapters \citep[e.g.][]{Jarvis2014,McAlpine2014,Takahashi2014,Bacon2014}, deep fields covered with the LSST will be prime candidates for undertaking a deep radio-continuum survey. Such data will allow accurate photometric redshifts, and therefore the $N(z)$ of the radio sources to be determined to a flux-density limit much fainter than possible over the wider area. Using these data will act as a benchmark for the assumptions needed for fully exploiting the wide-area surveys. The key issue is that the cosmic volume surveyed should be representative and not suffer significantly from sample variance, which could act to skew the redshift distribution if large-scale structures dominate certain redshift slices.
Moreover, such deep-fields provide us with the necessary information to understand how the radio sources trace the dark-matter distribution as we move from the single-halo term to the two-halo term at a few Mpcs.

In order to reach the requisite depth, to gain insights into the source population as a function of redshift from morphological classification, and to conduct the most efficient deep survey, SKA1-MID would be the facility of choice for this tier. This is because SKA1-SUR has a field-of-view much larger than the individual deep-drilling fields that will be targeted by the LSST, which are of order 10~deg$^2$. Furthermore, the  resolution required, at the relatively low frequency that such a survey needs to be conducted to reach the requisite source density/depth, in order to classify objects morphologically, would be extremely difficult to achieve with SKA1-SUR \citep[for details see ][]{Jarvis2014}.

\subsection{Medium-deep}

\noindent In practice, it will be necessary to add to the deep surveys to overcome sample variance, particular for the most highly-biased populations at moderate redshifts. However, the key driver for a medium-deep survey with SKA1 is to undertake the first wide-area weak lensing survey at radio wavelengths.
As detailed in \cite{Brown2014},  a survey covering $\sim 5000$~deg$^2$ to $\sim 1~\mu$Jy rms in Band 2, will provide excellent constraints on the power spectrum that are very complementary to weak-lensing surveys being conducted at optical wavelengths on a similar timescale.

\subsection{Wide}

\noindent The key survey for measuring the largest scales in the Universe is obviously a survey that can cover as much cosmological volume as possible. First, given that distant radio sources can be found in relatively shallow surveys due to their extreme brightness, then area is initially more important than extreme depth, and conducting an all-sky survey very early in the expansion up to SKA1 would be extremely beneficial for cosmology applications. 

In terms of which element of SKA1 is the best for such a survey, then the question is more open than for the deep and medium-deep tiers, as the areal coverage benefits from the large field-of-view offered by SKA1-SUR, allowing in principle the survey to be conducted on a shorter timescale. However, the resolution of $\sim 1$~arcsec at 1~GHz for SKA1-SUR precludes the science that requires the morphological characterisation of radio sources \citep[see e.g.][]{McAlpine2014, Makhatini2014}, which is needed to exploit the multi-tracer technique and  overcome the inhibiting effects of cosmic variance on the largest scales (Section~\ref{sec:powerspectrum}), or the different redshift distributions of different populations. Given that the cosmic variance is the dominant source of uncertainty on measurements of the power spectrum on the largest scales, then this is a crucial issue.

Although the large scales can be probed efficiently by SKA1-SUR, the additional benefits of high-resolution imaging, which pave the way for using the multi-tracer method using SKA1-MID, provide a unique probe of the largest scales. Given that SKA1-MID is required to conduct the weak lensing survey, and the expansion to all sky for SKA1-MID does not present a large increase in survey time compared to SKA1-SUR (a SKA1-MID survey can be carried out at lower frequency, $<1$~GHz, thus increasing the size of the primary beam, while still retaining the requisite resolution for morphological characterisation of the radio sources), then it would make sense to conduct even the widest of surveys with SKA1-MID.

\section{Towards SKA2}

\noindent
As we move towards SKA2 then the ability to conduct a world-leading weak lensing survey with completely different systematics to any optical survey becomes very compelling. A source density of $\sim 75$ sources per arcmin$^2$ over $3\pi$ steradians is within reach (assuming a detection threshold of  $S_{\rm 1~GHz} \ge 100$\,nJy), eclipsing the source density from current surveys, such as the Dark Energy Survey \citep[e.g.][]{Banerji2014}, and is higher density than {\em Euclid} will have. Furthermore, the PSF at radio wavelength is analytically predictable, and the additional information from polarisation (and H{\sc i} rotation curves) means that the  major source of uncertainty from intrinsic alignments may be overcome.
Thus, SKA continuum surveys will turn radio 
observations into a central probe of cosmological research in the coming decades. As shown above, SKA1 may already lead to transformational findings.

\section{Acknowledgments}
MJJ and MGS acknowledges support by the South African Square Kilometre Array
Project and the South African National Research Foundation. MGS also acknowledges support from FCT grant
PTDC/FIS-AST/2194/2012. DB is supported by UK Science and Technology Facilities Council, grant ST/K00090X/1. MLB is supported by an ERC Starting Grant (Grant no. 280127) and by a STFC Advanced/Halliday fellowship. AR is supported by the Templeton Foundation.
Part of the research described in this paper was carried out at the Jet Propulsion Laboratory, California Institute of Technology, under a contract with the National Aeronautics and Space Administration.

\bibliographystyle{apj_long_etal}
\bibliography{matts}

\begin{thebibliography}{49}
\expandafter\ifx\csname natexlab\endcsname\relax\def\natexlab#1{#1}\fi

\bibitem[{{Bacon} {et~al.}(2015){Bacon}, {Bridle}, {Brown}, {Bull}, {Camera},
  {Fender}, {Jarvis}, {Jackson}, {Kirk}, {Mann}, {McKean}, \&
  {Newman}}]{Bacon2014}
{Bacon}, D. et~al., . 2015, in PoS, Vol.~1, Advancing Astrophysics with the SKA

\bibitem[{{Banerji} {et~al.}(2014){Banerji}, {Jouvel}, {Lin}, {McMahon},
  {Lahav}, {Castander}, {Abdalla}, {Bertin}, {Bosman}, {Carnero}, {Carrasco
  Kind}, {da Costa}, {Gerdes}, {Gschwend}, {Lima}, {Maia}, {Merson}, {Miller},
  {Ogando}, {Pellegrini}, {Reed}, {Saglia}, {Sanchez}, {Annis}, {Bernstein},
  {Bernstein}, {Bernstein}, {Capozzi}, {Childress}, {Cunha}, {Davis}, {DePoy},
  {Desai}, {Diehl}, {Doel}, {Findlay}, {Finley}, {Flaugher}, {Frieman},
  {Gaztanaga}, {Glazebrook}, {Gonzalez-Fernandez}, {Gonzalez-Solares},
  {Honscheid}, {Irwin}, {Jarvis}, {Kim}, {Koposov}, {Kuehn}, {Kupcu-Yoldas},
  {Lagattuta}, {Lewis}, {Lidman}, {Makler}, {Marriner}, {Marshall}, {Miquel},
  {Mohr}, {Neilsen}, {Peoples}, {Sako}, {Sanchez}, {Scarpine}, {Schindler},
  {Schubnell}, {Sevilla}, {Sharp}, {Soares-Santos}, {Swanson}, {Tarle},
  {Thaler}, {Tucker}, {Uddin}, {Wechsler}, {Wester}, {Yuan}, \&
  {Zuntz}}]{Banerji2014}
{Banerji}, M. et~al., . 2014, ArXiv.1407.3801

\bibitem[{{Blake} {et~al.}(2012){Blake}, {Brough}, {Colless}, {Contreras},
  {Couch}, {Croom}, {Croton}, {Davis}, {Drinkwater}, {Forster}, {Gilbank},
  {Gladders}, {Glazebrook}, {Jelliffe}, {Jurek}, {Li}, {Madore}, {Martin},
  {Pimbblet}, {Poole}, {Pracy}, {Sharp}, {Wisnioski}, {Woods}, {Wyder}, \&
  {Yee}}]{Blake2012}
{Blake}, C. et~al., . 2012, \mnras, 425, 405

\bibitem[{{Brown} {et~al.}(2015){Brown}, {Bacon}, {Camera}, {Harrison},
  {Joachimi}, {Metcalfe}, {Pourtsidou}, {Takahashi}, {Zuntz}, {Abdalla},
  {Bridle}, {Jarvis}, {Kitching}, {Miller}, \& {Patel}}]{Brown2014}
{Brown}, M. et~al., . 2015, in PoS, Vol.~1, Advancing Astrophysics with the SKA

\bibitem[{{Bull} {et~al.}(2015){Bull}, {Camera}, {Raccanelli}, {Blake},
  {Ferreira}, {Santos}, \& {Schwarz}}]{Bull2014}
{Bull}, P., {Camera}, S., {Raccanelli}, A., {Blake}, C., {Ferreira}, P.,
  {Santos}, M., \& {Schwarz}, D. 2015, in PoS, Vol.~1, Advancing Astrophysics
  with the SKA

\bibitem[{{Camera} {et~al.}(2012){Camera}, {Santos}, {Bacon}, {Jarvis},
  {McAlpine}, {Norris}, {Raccanelli}, \& {R{\"o}ttgering}}]{Camera2012}
{Camera}, S., {Santos}, M.~G., {Bacon}, D.~J., {Jarvis}, M.~J., {McAlpine}, K.,
  {Norris}, R.~P., {Raccanelli}, A., \& {R{\"o}ttgering}, H. 2012, \mnras, 427,
  2079

\bibitem[{{Cirasuolo} {et~al.}(2012){Cirasuolo}, {Afonso}, {Bender},
  {Bonifacio}, {Evans}, {Kaper}, {Oliva}, {Vanzi}, {Abreu}, {Atad-Ettedgui},
  {Babusiaux}, {Bauer}, {Best}, {Bezawada}, {Bryson}, {Cabral}, {Caputi},
  {Centrone}, {Chemla}, {Cimatti}, {Cioni}, {Clementini}, {Coelho}, {Daddi},
  {Dunlop}, {Feltzing}, {Ferguson}, {Flores}, {Fontana}, {Fynbo}, {Garilli},
  {Glauser}, {Guinouard}, {Hammer}, {Hastings}, {Hess}, {Ivison}, {Jagourel},
  {Jarvis}, {Kauffman}, {Lawrence}, {Lee}, {Li Causi}, {Lilly}, {Lorenzetti},
  {Maiolino}, {Mannucci}, {McLure}, {Minniti}, {Montgomery}, {Muschielok},
  {Nandra}, {Navarro}, {Norberg}, {Origlia}, {Padilla}, {Peacock}, {Pedicini},
  {Pentericci}, {Pragt}, {Puech}, {Randich}, {Renzini}, {Ryde}, {Rodrigues},
  {Royer}, {Saglia}, {S{\'a}nchez}, {Schnetler}, {Sobral}, {Speziali}, {Todd},
  {Tolstoy}, {Torres}, {Venema}, {Vitali}, {Wegner}, {Wells}, {Wild}, \&
  {Wright}}]{CirasuoloMOONS2012}
{Cirasuolo}, M. et~al., . 2012, in Society of Photo-Optical Instrumentation
  Engineers (SPIE) Conference Series, Vol. 8446, Society of Photo-Optical
  Instrumentation Engineers (SPIE) Conference Series

\bibitem[{{Cirasuolo} {et~al.}(2014){Cirasuolo}, {Afonso}, {Carollo}, {Flores},
  {Maiolino}, {Oliva}, {Paltani}, {Vanzi}, {Evans}, {Abreu}, {Atkinson},
  {Babusiaux}, {Beard}, {Bauer}, {Bellazzini}, {Bender}, {Best}, {Bezawada},
  {Bonifacio}, {Bragaglia}, {Bryson}, {Busher}, {Cabral}, {Caputi}, {Centrone},
  {Chemla}, {Cimatti}, {Cioni}, {Clementini}, {Coelho}, {Crnojevic}, {Daddi},
  {Dunlop}, {Eales}, {Feltzing}, {Ferguson}, {Fisher}, {Fontana}, {Fynbo},
  {Garilli}, {Gilmore}, {Glauser}, {Guinouard}, {Hammer}, {Hastings}, {Hess},
  {Ivison}, {Jagourel}, {Jarvis}, {Kaper}, {Kauffman}, {Kitching}, {Lawrence},
  {Lee}, {Lemasle}, {Licausi}, {Lilly}, {Lorenzetti}, {Lunney}, {Maiolino},
  {Mannucci}, {McLure}, {Minniti}, {Montgomery}, {Muschielok}, {Nandra},
  {Navarro}, {Norberg}, {Oliver}, {Origlia}, {Padilla}, {Peacock}, {Pedichini},
  {Peng}, {Pentericci}, {Pragt}, {Puech}, {Randich}, {Rees}, {Renzini}, {Ryde},
  {Rodrigues}, {Roseboom}, {Royer}, {Saglia}, {Sanchez}, {Schiavon},
  {Schnetler}, {Sobral}, {Speziali}, {Sun}, {Stuik}, {Taylor}, {Taylor},
  {Todd}, {Tolstoy}, {Torres}, {Tosi}, {Vanzella}, {Venema}, {Vitali},
  {Wegner}, {Wells}, {Wild}, {Wright}, {Zamorani}, \& {Zoccali}}]{MOONS2}
{Cirasuolo}, M. et~al., . 2014, in Society of Photo-Optical Instrumentation
  Engineers (SPIE) Conference Series, Vol. 9147, Society of Photo-Optical
  Instrumentation Engineers (SPIE) Conference Series, 0

\bibitem[{{Clarkson} {et~al.}(2008){Clarkson}, {Bassett}, \&
  {Lu}}]{Clarkson2008}
{Clarkson}, C., {Bassett}, B., \& {Lu}, T.~H.-C. 2008, Physical Review Letters,
  101, 011301

\bibitem[{{Condon} {et~al.}(2012){Condon}, {Cotton}, {Fomalont}, {Kellermann},
  {Miller}, {Perley}, {Scott}, {Vernstrom}, \& {Wall}}]{Condon2012}
{Condon}, J.~J., {Cotton}, W.~D., {Fomalont}, E.~B., {Kellermann}, K.~I.,
  {Miller}, N., {Perley}, R.~A., {Scott}, D., {Vernstrom}, T., \& {Wall}, J.~V.
  2012, \apj, 758, 23

\bibitem[{{Condon} {et~al.}(1998){Condon}, {Cotton}, {Greisen}, {Yin},
  {Perley}, {Taylor}, \& {Broderick}}]{Condon1998}
{Condon}, J.~J., {Cotton}, W.~D., {Greisen}, E.~W., {Yin}, Q.~F., {Perley},
  R.~A., {Taylor}, G.~B., \& {Broderick}, J.~J. 1998, \aj, 115, 1693

\bibitem[{{Copi} {et~al.}(2013{\natexlab{a}}){Copi}, {Huterer}, {Schwarz}, \&
  {Starkman}}]{Copi2013a}
{Copi}, C.~J., {Huterer}, D., {Schwarz}, D.~J., \& {Starkman}, G.~D.
  2013{\natexlab{a}}, ArXiv.1310.3831

\bibitem[{{Copi} {et~al.}(2013{\natexlab{b}}){Copi}, {Huterer}, {Schwarz}, \&
  {Starkman}}]{Copi2013b}
---. 2013{\natexlab{b}}, ArXiv.1311.4562

\bibitem[{{Das} {et~al.}(2014){Das}, {Louis}, {Nolta}, {Addison},
  {Battistelli}, {Bond}, {Calabrese}, {Crichton}, {Devlin}, {Dicker},
  {Dunkley}, {D{\"u}nner}, {Fowler}, {Gralla}, {Hajian}, {Halpern},
  {Hasselfield}, {Hilton}, {Hincks}, {Hlozek}, {Huffenberger}, {Hughes},
  {Irwin}, {Kosowsky}, {Lupton}, {Marriage}, {Marsden}, {Menanteau}, {Moodley},
  {Niemack}, {Page}, {Partridge}, {Reese}, {Schmitt}, {Sehgal}, {Sherwin},
  {Sievers}, {Spergel}, {Staggs}, {Swetz}, {Switzer}, {Thornton}, {Trac}, \&
  {Wollack}}]{ACTLens}
{Das}, S. et~al., . 2014, \jcap, 4, 14

\bibitem[{{de Jong} {et~al.}(2013){de Jong}, {Kuijken}, {Applegate}, {Begeman},
  {Belikov}, {Blake}, {Bout}, {Boxhoorn}, {Buddelmeijer}, {Buddendiek},
  {Cacciato}, {Capaccioli}, {Choi}, {Cordes}, {Covone}, {Dall'Ora}, {Edge},
  {Erben}, {Franse}, {Getman}, {Grado}, {Harnois-Deraps}, {Helmich},
  {Herbonnet}, {Heymans}, {Hildebrandt}, {Hoekstra}, {Huang}, {Irisarri},
  {Joachimi}, {K{\"o}hlinger}, {Kitching}, {La Barbera}, {Lacerda},
  {McFarland}, {Miller}, {Nakajima}, {Napolitano}, {Paolillo}, {Peacock},
  {Pila-Diez}, {Puddu}, {Radovich}, {Rifatto}, {Schneider}, {Schrabback},
  {Sifon}, {Sikkema}, {Simon}, {Sutherland}, {Tudorica}, {Valentijn}, {van der
  Burg}, {van Uitert}, {van Waerbeke}, {Velander}, {Kleijn}, {Viola}, \&
  {Vriend}}]{KIDS}
{de Jong}, J.~T.~A. et~al., . 2013, The Messenger, 154, 44

\bibitem[{{de Jong} {et~al.}(2012){de Jong}, {Bellido-Tirado}, {Chiappini},
  {Depagne}, {Haynes}, {Johl}, {Schnurr}, {Schwope}, {Walcher}, {Dionies},
  {Haynes}, {Kelz}, {Kitaura}, {Lamer}, {Minchev}, {M{\"u}ller}, {Nuza},
  {Olaya}, {Piffl}, {Popow}, {Steinmetz}, {Ural}, {Williams}, {Winkler},
  {Wisotzki}, {Ansorge}, {Banerji}, {Gonzalez Solares}, {Irwin}, {Kennicutt},
  {King}, {McMahon}, {Koposov}, {Parry}, {Sun}, {Walton}, {Finger}, {Iwert},
  {Krumpe}, {Lizon}, {Vincenzo}, {Amans}, {Bonifacio}, {Cohen}, {Francois},
  {Jagourel}, {Mignot}, {Royer}, {Sartoretti}, {Bender}, {Grupp}, {Hess},
  {Lang-Bardl}, {Muschielok}, {B{\"o}hringer}, {Boller}, {Bongiorno}, {Brusa},
  {Dwelly}, {Merloni}, {Nandra}, {Salvato}, {Pragt}, {Navarro}, {Gerlofsma},
  {Roelfsema}, {Dalton}, {Middleton}, {Tosh}, {Boeche}, {Caffau}, {Christlieb},
  {Grebel}, {Hansen}, {Koch}, {Ludwig}, {Quirrenbach}, {Sbordone}, {Seifert},
  {Thimm}, {Trifonov}, {Helmi}, {Trager}, {Feltzing}, {Korn}, \&
  {Boland}}]{4MOST}
{de Jong}, R.~S. et~al., . 2012, in Society of Photo-Optical Instrumentation
  Engineers (SPIE) Conference Series, Vol. 8446, Society of Photo-Optical
  Instrumentation Engineers (SPIE) Conference Series

\bibitem[{{Edge} {et~al.}(2013){Edge}, {Sutherland}, {Kuijken}, {Driver},
  {McMahon}, {Eales}, \& {Emerson}}]{VIKING}
{Edge}, A., {Sutherland}, W., {Kuijken}, K., {Driver}, S., {McMahon}, R.,
  {Eales}, S., \& {Emerson}, J.~P. 2013, The Messenger, 154, 32

\bibitem[{{Ferramacho} {et~al.}(2014){Ferramacho}, {Santos}, {Jarvis}, \&
  {Camera}}]{Ferramacho2014}
{Ferramacho}, L.~D., {Santos}, M.~G., {Jarvis}, M.~J., \& {Camera}, S. 2014,
  \mnras, 442, 2511

\bibitem[{{Geach} {et~al.}(2013){Geach}, {Hickox}, {Bleem}, {Brodwin},
  {Holder}, {Aird}, {Benson}, {Bhattacharya}, {Carlstrom}, {Chang}, {Cho},
  {Crawford}, {Crites}, {de Haan}, {Dobbs}, {Dudley}, {George}, {Hainline},
  {Halverson}, {Holzapfel}, {Hoover}, {Hou}, {Hrubes}, {Keisler}, {Knox},
  {Lee}, {Leitch}, {Lueker}, {Luong-Van}, {Marrone}, {McMahon}, {Mehl},
  {Meyer}, {Millea}, {Mohr}, {Montroy}, {Myers}, {Padin}, {Plagge}, {Pryke},
  {Reichardt}, {Ruhl}, {Sayre}, {Schaffer}, {Shaw}, {Shirokoff}, {Spieler},
  {Staniszewski}, {Stark}, {Story}, {van Engelen}, {Vanderlinde}, {Vieira},
  {Williamson}, \& {Zahn}}]{Geach2013}
{Geach}, J.~E. et~al., . 2013, \apjl, 776, L41

\bibitem[{{Giannantonio} {et~al.}(2012){Giannantonio}, {Crittenden}, {Nichol},
  \& {Ross}}]{Giannantonio2012}
{Giannantonio}, T., {Crittenden}, R., {Nichol}, R., \& {Ross}, A.~J. 2012,
  \mnras, 426, 2581

\bibitem[{{Giannantonio} {et~al.}(2008){Giannantonio}, {Scranton},
  {Crittenden}, {Nichol}, {Boughn}, {Myers}, \& {Richards}}]{Giannantonio2008}
{Giannantonio}, T., {Scranton}, R., {Crittenden}, R.~G., {Nichol}, R.~C.,
  {Boughn}, S.~P., {Myers}, A.~D., \& {Richards}, G.~T. 2008, \prd, 77, 123520

\bibitem[{{Heymans} {et~al.}(2012){Heymans}, {Van Waerbeke}, {Miller}, {Erben},
  {Hildebrandt}, {Hoekstra}, {Kitching}, {Mellier}, {Simon}, {Bonnett},
  {Coupon}, {Fu}, {Harnois D{\'e}raps}, {Hudson}, {Kilbinger}, {Kuijken},
  {Rowe}, {Schrabback}, {Semboloni}, {van Uitert}, {Vafaei}, \&
  {Velander}}]{Heymans2012}
{Heymans}, C. et~al., . 2012, \mnras, 427, 146

\bibitem[{{Jarvis} {et~al.}(2013){Jarvis}, {Bonfield}, {Bruce}, {Geach},
  {McAlpine}, {McLure}, {Gonz{\'a}lez-Solares}, {Irwin}, {Lewis}, {Yoldas},
  {Andreon}, {Cross}, {Emerson}, {Dalton}, {Dunlop}, {Hodgkin}, {Le},
  {Karouzos}, {Meisenheimer}, {Oliver}, {Rawlings}, {Simpson}, {Smail},
  {Smith}, {Sullivan}, {Sutherland}, {White}, \& {Zwart}}]{Jarvis2013}
{Jarvis}, M.~J. et~al., . 2013, \mnras, 428, 1281

\bibitem[{{Jarvis} {et~al.}(2015){Jarvis}, {Seymour}, {Afonso}, {Best},
  {Beswick}, {Heywood}, {Huynh}, {Murphy}, {Prandoni}, {Simpson}, {Vaccari}, \&
  {White}}]{Jarvis2014}
{Jarvis}, M.~J. et~al., . 2015, in PoS, Vol.~1, Advancing Astrophysics with the
  SKA

\bibitem[{{Kitching} {et~al.}(2015){Kitching}, {Bacon}, {Brown}, {Bull},
  {McEwen}, {Oguri}, {Scaramella}, {Takahashi}, {Wu}, \&
  {Yamauchi}}]{Kitching2014}
{Kitching}, T. et~al., . 2015, in PoS, Vol.~1, Advancing Astrophysics with the
  SKA

\bibitem[{{Li} \& {Barrow}(2009)}]{LiBarrow2009}
{Li}, B. \& {Barrow}, J.~D. 2009, \prd, 79, 103521

\bibitem[{{Lindsay} {et~al.}(2014{\natexlab{a}}){Lindsay}, {Jarvis}, \&
  {McAlpine}}]{Lindsay2014b}
{Lindsay}, S.~N., {Jarvis}, M.~J., \& {McAlpine}, K. 2014{\natexlab{a}},
  \mnras, 440, 2322

\bibitem[{{Lindsay} {et~al.}(2014{\natexlab{b}}){Lindsay}, {Jarvis}, {Santos},
  {Brown}, {Croom}, {Driver}, {Hopkins}, {Liske}, {Loveday}, {Norberg}, \&
  {Robotham}}]{Lindsay2014a}
{Lindsay}, S.~N. et~al., . 2014{\natexlab{b}}, \mnras, 440, 1527

\bibitem[{{Makhatini} {et~al.}(2015){Makhatini}, {Smirnov}, {Jarvis}, \&
  {Heywood}}]{Makhatini2014}
{Makhatini}, S., {Smirnov}, O., {Jarvis}, M., \& {Heywood}, I. 2015, in PoS,
  Vol.~1, Advancing Astrophysics with the SKA

\bibitem[{{McAlpine} {et~al.}(2015){McAlpine}, {Prandoni}, {Jarvis}, {Seymour},
  {Padovani}, {Best}, {Simpson}, {Guidetti}, {Murphy}, {Huynh}, {Vaccari}, \&
  {White}}]{McAlpine2014}
{McAlpine}, K. et~al., . 2015, in PoS, Vol.~1, Advancing Astrophysics with the
  SKA

\bibitem[{{McCracken} {et~al.}(2012){McCracken}, {Milvang-Jensen}, {Dunlop},
  {Franx}, {Fynbo}, {Le F{\`e}vre}, {Holt}, {Caputi}, {Goranova}, {Buitrago},
  {Emerson}, {Freudling}, {Hudelot}, {L{\'o}pez-Sanjuan}, {Magnard}, {Mellier},
  {M{\o}ller}, {Nilsson}, {Sutherland}, {Tasca}, \& {Zabl}}]{McCracken2012}
{McCracken}, H.~J. et~al., . 2012, \aap, 544, A156

\bibitem[{{Nolta} {et~al.}(2004){Nolta}, {Wright}, {Page}, {Bennett},
  {Halpern}, {Hinshaw}, {Jarosik}, {Kogut}, {Limon}, {Meyer}, {Spergel},
  {Tucker}, \& {Wollack}}]{Nolta2004}
{Nolta}, M.~R. et~al., . 2004, \apj, 608, 10

\bibitem[{{Patel} {et~al.}(2015){Patel}, {Harrison}, {Makhatini}, {Abdalla},
  {Bacon}, {Brown}, {Jarvis}, \& {Smirnov}}]{Patel2014}
{Patel}, P., {Harrison}, I., {Makhatini}, S., {Abdalla}, F., {Bacon}, D.,
  {Brown}, M., {Jarvis}, M., \& {Smirnov}, O. 2015, in PoS, Vol.~1, Advancing
  Astrophysics with the SKA

\bibitem[{{Planck Collaboration} {et~al.}(2013{\natexlab{a}}){Planck
  Collaboration}, {Ade}, {Aghanim}, {Armitage-Caplan}, {Arnaud}, {Ashdown},
  {Atrio-Barandela}, {Aumont}, {Baccigalupi}, {Banday}, \& et~al.}]{PlanckLens}
{Planck Collaboration} et~al., . 2013{\natexlab{a}}, ArXiv.1303.5077

\bibitem[{{Planck Collaboration} {et~al.}(2013{\natexlab{b}}){Planck
  Collaboration}, {Ade}, {Aghanim}, {Armitage-Caplan}, {Arnaud}, {Ashdown},
  {Atrio-Barandela}, {Aumont}, {Baccigalupi}, {Banday}, \&
  et~al.}]{PlanckIsotropy}
---. 2013{\natexlab{b}}, ArXiv.1303.5083

\bibitem[{{Planck Collaboration} {et~al.}(2013{\natexlab{c}}){Planck
  Collaboration}, {Ade}, {Aghanim}, {Armitage-Caplan}, {Arnaud}, {Ashdown},
  {Atrio-Barandela}, {Aumont}, {Baccigalupi}, {Banday}, \& et~al.}]{PlanckNG}
---. 2013{\natexlab{c}}, ArXiv.1303.5084

\bibitem[{{Raccanelli} {et~al.}(2008){Raccanelli}, {Bonaldi}, {Negrello},
  {Matarrese}, {Tormen}, \& {de Zotti}}]{Raccanelli2008}
{Raccanelli}, A., {Bonaldi}, A., {Negrello}, M., {Matarrese}, S., {Tormen}, G.,
  \& {de Zotti}, G. 2008, \mnras, 386, 2161

\bibitem[{{Raccanelli} {et~al.}(2014){Raccanelli}, {Dor{\'e}}, {Bacon},
  {Maartens}, {Santos}, {Camera}, {Davis}, {Drinkwater}, {Jarvis}, {Norris}, \&
  {Parkinson}}]{Raccanelli2014}
{Raccanelli}, A. et~al., . 2014, ArXiv.1406.0010

\bibitem[{{Raccanelli} {et~al.}(2012){Raccanelli}, {Zhao}, {Bacon}, {Jarvis},
  {Percival}, {Norris}, {R{\"o}ttgering}, {Abdalla}, {Cress}, {Kubwimana},
  {Lindsay}, {Nichol}, {Santos}, \& {Schwarz}}]{Raccanelli2012}
---. 2012, \mnras, 424, 801

\bibitem[{{Sachs} \& {Wolfe}(1967)}]{SachsWolfe1967}
{Sachs}, R.~K. \& {Wolfe}, A.~M. 1967, \apj, 147, 73

\bibitem[{{Schwarz} {et~al.}(2015){Schwarz}, {Bacon}, {Chen}, {Clarkson},
  {Huterer}, {Kunz}, {Maartens}, {Raccanelli}, {Rubart}, \&
  {Stack}}]{Schwarz2014}
{Schwarz}, D. et~al., . 2015, in PoS, Vol.~1, Advancing Astrophysics with the
  SKA

\bibitem[{{Seljak}(2009)}]{Seljak2009}
{Seljak}, U. 2009, Physical Review Letters, 102, 021302

\bibitem[{{Sherwin} {et~al.}(2012){Sherwin}, {Das}, {Hajian}, {Addison},
  {Bond}, {Crichton}, {Devlin}, {Dunkley}, {Gralla}, {Halpern}, {Hill},
  {Hincks}, {Hughes}, {Huffenberger}, {Hlozek}, {Kosowsky}, {Louis},
  {Marriage}, {Marsden}, {Menanteau}, {Moodley}, {Niemack}, {Page}, {Reese},
  {Sehgal}, {Sievers}, {Sif{\'o}n}, {Spergel}, {Staggs}, {Switzer}, \&
  {Wollack}}]{Sherwin2012}
{Sherwin}, B.~D. et~al., . 2012, \prd, 86, 083006

\bibitem[{{Takada} {et~al.}(2014){Takada}, {Ellis}, {Chiba}, {Greene},
  {Aihara}, {Arimoto}, {Bundy}, {Cohen}, {Dor{\'e}}, {Graves}, {Gunn},
  {Heckman}, {Hirata}, {Ho}, {Kneib}, {F{\`e}vre}, {Lin}, {More}, {Murayama},
  {Nagao}, {Ouchi}, {Seiffert}, {Silverman}, {Sodr{\'e}}, {Spergel}, {Strauss},
  {Sugai}, {Suto}, {Takami}, \& {Wyse}}]{TakadaPFS2014}
{Takada}, M. et~al., . 2014, \pasj, 66, 1

\bibitem[{{Takahashi} {et~al.}(2015){Takahashi}, {Brown}, {Burigana},
  {Jackson}, {Jarvis}, {Kitching}, {Kneib}, {Oguri}, {Prunet}, {Shan},
  {Starck}, \& {Yamauchi}}]{Takahashi2014}
{Takahashi}, K. et~al., . 2015, in PoS, Vol.~1, Advancing Astrophysics with the
  SKA

\bibitem[{{Velten} \& {Schwarz}(2011)}]{Velten2011}
{Velten}, H. \& {Schwarz}, D.~J. 2011, \jcap, 9, 16

\bibitem[{{Wilman} {et~al.}(2010){Wilman}, {Jarvis}, {Mauch}, {Rawlings}, \&
  {Hickey}}]{Wilman2010}
{Wilman}, R.~J., {Jarvis}, M.~J., {Mauch}, T., {Rawlings}, S., \& {Hickey}, S.
  2010, \mnras, 405, 447

\bibitem[{{Wilman} {et~al.}(2008){Wilman}, {Miller}, {Jarvis}, {Mauch},
  {Levrier}, {Abdalla}, {Rawlings}, {Kl{\"o}ckner}, {Obreschkow}, {Olteanu}, \&
  {Young}}]{Wilman2008}
{Wilman}, R.~J. et~al., . 2008, \mnras, 388, 1335

\bibitem[{{Xia} {et~al.}(2010){Xia}, {Viel}, {Baccigalupi}, {De Zotti},
  {Matarrese}, \& {Verde}}]{Xia2010}
{Xia}, J.-Q., {Viel}, M., {Baccigalupi}, C., {De Zotti}, G., {Matarrese}, S.,
  \& {Verde}, L. 2010, \apjl, 717, L17

\end{thebibliography}

\end{document}